# The use of Extraterrestrial Resources to Facilitate Space Science and Exploration

**Ian A. Crawford[1], Martin Elvis[2] and James Carpenter[3]**

[1]Department of Earth and Planetary Sciences, Birkbeck College, London, UK;
[2]Harvard-Smithsonian Center for Astrophysics, Cambridge, Massachusetts, USA;
[3]ESA-ESTEC, Noordwijk, The Netherlands.

--------

**Meeting Report:** Ian Crawford, Martin Elvis and James Carpenter summarize a Royal Astronomical Society Specialist Discussion Meeting which examined how science will benefit from the use of extraterrestrial resources.

To-date, all human economic activity has depended on the resources of a single planet, and it has long been recognized that developments in space exploration could in principle open our closed planetary economy to external resources of energy and raw materials. Recently, there has been renewed interest in these possibilities, with several private companies established with the stated aim of exploiting extraterrestrial resources. Space science and exploration are among the potential beneficiaries of space resources because their use may permit the construction and operation of scientific facilities in space that will be unaffordable if all the required material and energy resources have to be lifted out of Earth's gravity. Examples may include the next generation of large space telescopes, sample return missions to the outer Solar System, and human research stations on the Moon and Mars. These potential scientific benefits of extraterrestrial resource utilisation were the topic of a Specialist Discussion meeting held at Burlington House on 8 April 2016.

**Martin Elvis** (Harvard Smithsonian Center for Astrophysics) got the meeting underway by asking "*What can space resources do for astronomy?*" Martin made the point that in order to observe the distant Universe, or make detailed studies of planets around other stars, astronomers will require larger telescopes in space across the electromagnetic spectrum, but these are currently unaffordable. The aging 'Great Observatories' (Chandra, HST, and Spitzer) will soon come to an end, and although the infra-red capabilities of Spitzer will be taken over, and enhanced, by the JWST from 2018, the cost of just this one telescope will have consumed almost two decade's worth of NASA's available funding for large astrophysics missions. The prospects for providing simultaneous coverage at other wavelengths with comparable instruments, or building the much larger instruments that will be required, for example, to resolve surface features on Earth-sized exoplanets (Fig. 1), seem bleak based on existing funding models. As the cost of large space telescopes (and also of planetary missions) increases much faster than economic growth, a 'funding wall' will soon be encountered, effectively ending the growth in space astronomy to which we have become accustomed. Martin therefore argued that a change in paradigm is required. The astronomy community should exploit synergies with the emerging 'new space' entrepreneurs who are seeking to reduce launch costs, develop space tourism, and mine the Moon and asteroids for profitable materials (e.g. Elvis, 2012). Within 5 years these commercial activities should cut mission costs significantly, and by 2030 will enable the next generation of large-scale astrophysics and planetary missions on a scale that the exploration of the Universe requires.

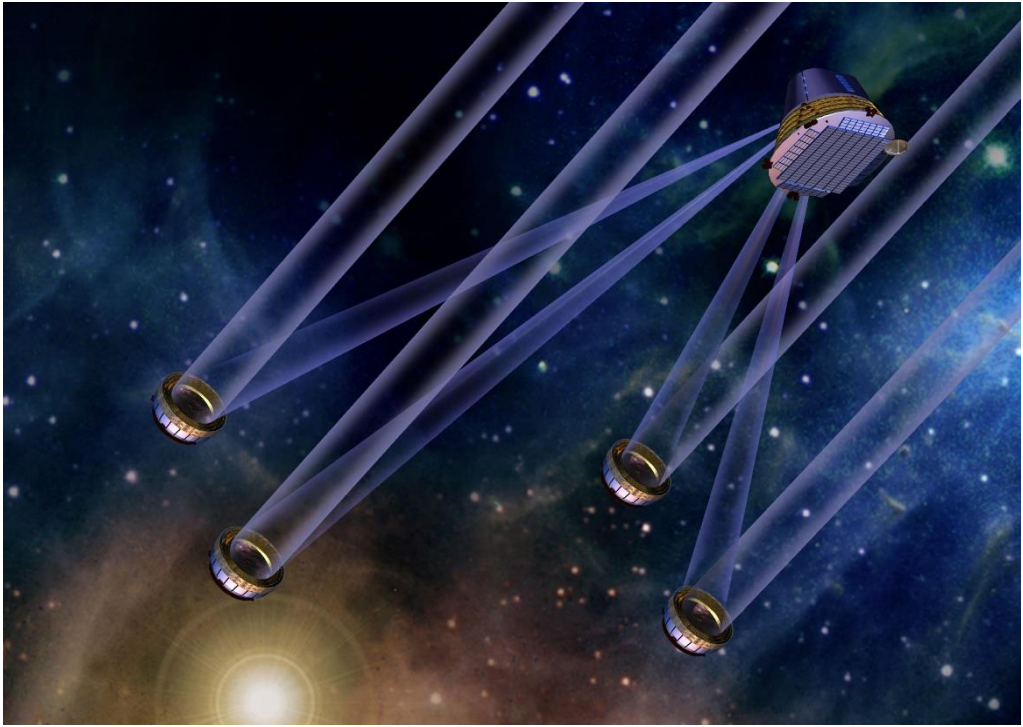

**Fig. 1**. Artist's conception of a large space-based optical/IR interferometer of the kind that may one day be required to study Earth-like planets orbiting other stars, and other astrophysical studies requiring large light grasp and very high angular resolution. Such instruments are unlikely to be affordable with current funding models, but may become so in the context of a future space economy built around the use of extraterrestrial resources (NASA/MPIA/T. Herbst).

Martin was followed by **Paul Spudis** (Lunar and Planetary Institute), who discussed "*The Moon as an enabling asset for spaceflight*." Paul argued that the Moon can act as a "gateway to the Universe" because its proximity, and resource-rich nature, will permit the development of a transportation infrastructure in cislunar space that will allow us to reach any point in the Solar System. The unique properties of the lunar poles (Fig. 2) are especially valuable in this context. At the poles, quasi-permanent, grazing incident sunlight will enable the generation of energy for constant operations, and also results in a benign surface thermal environment. Moreover, water ice near the poles can support human life, protect the crew from galactic and solar radiation, and serve as a medium for energy storage. Most importantly, the water may be converted into cryogenic oxygen and hydrogen, the most powerful chemical rocket propellant known. These energy and material resources will allow us to scientifically explore the Moon, and therefore learn about the history and evolution of the Earth-Moon system, the Solar System, and the wider Universe. In addition, lunar resources will allow us to develop an economic infrastructure in cislunar space where most of our satellite assets reside, including those devoted to observational astronomy (see Spudis, 2016). From a planetary science perspective, a cis-lunar transportation infrastructure based on lunar resources will permit sustainable human missions to Mars, and perhaps other planets, that are unlikely to be possible otherwise.

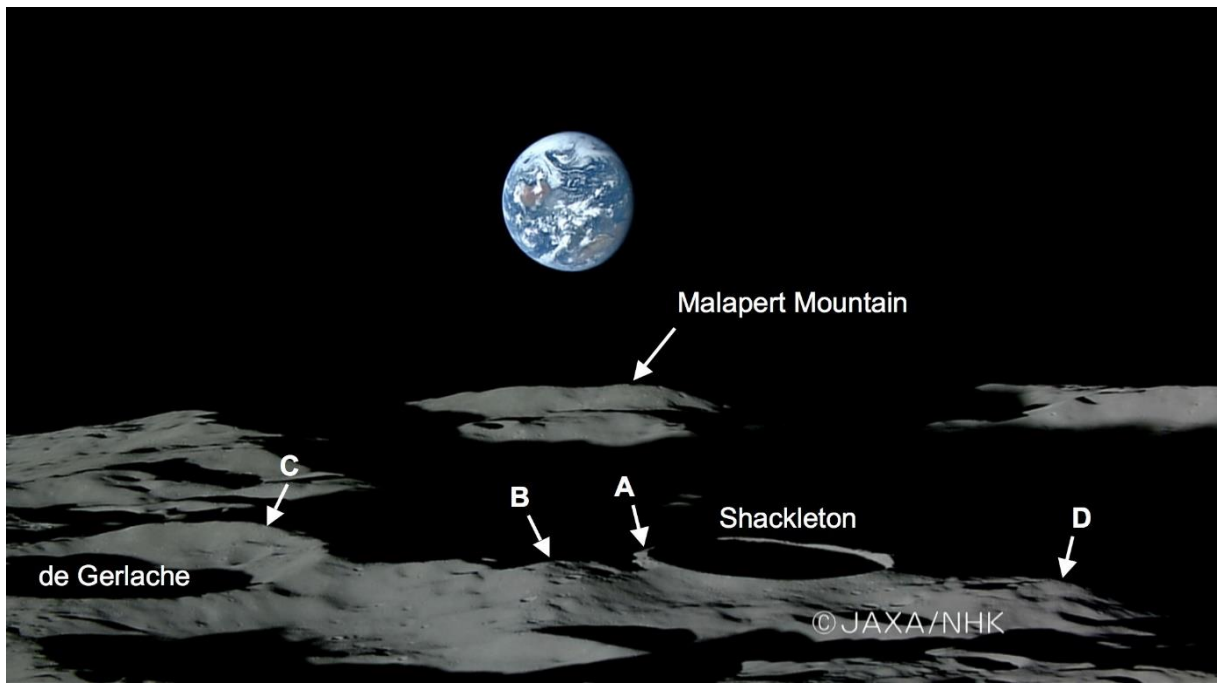

**Fig. 2**. The south pole of the Moon in the vicinity of the 20 km crater Shackleton, imaged by Japan's Kaguya spacecraft. The image shows four points (A-D) which receive sunlight more than 80% of a year. In contrast, the interiors of polar craters such as Shackleton are in permanent shadow and very cold (<40 K), and there is strong evidence that water ice has accumulated within them. Many speakers drew attention to the importance of this water as a source of $H_2$ and $O_2$ to facilitate future space exploration and development (JAXA/NHK/P. Spudis/Wikipedia).

The latter theme was picked up by **Olivier de Weck** (Massachusetts Institute of Technology) who presented a paper entitled "*Detour to the Moon: How lunar resources can save up to 68% of launch mass to Mars.*" This gave a detailed analysis of interplanetary supply chains taking account of *in situ* resource utilisation (ISRU) on planetary surfaces. Their results (reported in detail by Ishimatsu et al., 2015) indicate that lunar ISRU water production, and the use of aerocapture at Mars, can be important in reducing the mass that needs to be launched into Earth orbit for human Mars missions. Specifically, they find that a strategy using lunar resources may reduce the overall mass required to be launched into low Earth orbit for recurring Mars missions by 68% relative to NASA's Mars Design Reference Architecture 5.0 (NASA, 2009), even when including the mass of the ISRU infrastructures that will need to be pre-deployed on the Moon. Lunar ISRU becomes attractive for $O_2$ productivity levels above about 1.8 kg per year per kg of ISRU plant mass. Their work also indicates that use of liquid oxygen/liquid hydrogen propulsion will be more advantageous than some suggested alternatives (e.g. nuclear thermal propulsion) precisely because it is more compatible with ISRU-derived propellants.

Olivier was followed by **Phil Metzger** (University of Central Florida) who talked on "*Using ice for planetary surface propulsion: a strategic technology to initiate space industry*." Phil noted that water is the most strategic resource for initial space mining efforts. Potential customers include space agencies conducting scientific exploration, as well as commercial satellite companies wishing to boost spacecraft to higher orbits. The economic problem is that a substantial infrastructure will be needed in space to mine the water, electrolyze it, cryogenically store the resulting $H_2$ and $O_2$, and transfer them to other spacecraft. Phil argued that the start-up costs could be reduced by using water directly as a propellant through heating it with concentrated sunlight to form steam. A first step would be to develop these technologies for planetary science missions to icy bodies, and to apply them in a cislunar water economy later. For example, robotic missions to the ice caps of Mars, or the surface of Europa, could utilise steam propulsion for mobility. This would enable more efficient planetary exploration because the surface landers would be able to move between multiple sites of scientific interest without exhausting their propellant. Initial laboratory demonstrations and analyses, including detailed thermodynamic modelling, indicate that a 6U cubesat-sized spacecraft using steam propulsion could hop on the order of a kilometer on Mars and multiple kilometers on Europa or Pluto. Depending on the volume of the water tank, a 155 kg lander (i.e. Mars Phoenix class) could hop tens or even hundreds of kilometers on Europa.

The next talk was talk was given by **Jim Keravala** (Shackleton Energy Company) who spoke on "*Accelerating space science and exploration in partnership with commercial resource utilisation.*" Given the increasing world population, and ever rising demands for energy and raw materials, Jim argued that access to lunar resources will prove to be a cornerstone requirement for the continued development of the world economy as well as for a sustained future in space. The establishment of propellant depots, based on lunar-derived $H_2$ and $O_2$, at key locations in near-Earth space will enable reusable space transportation to become a viable proposition. This will require access to water sourced from the cold traps of polar lunar craters (Fig. 2) that will then provide the backbone infrastructure for growth of a space-based economy. Shackleton Energy (http://www.shackletonenergy.com/) was founded from the space, mining, energy and exploration sectors to meet this challenge as a fully private venture. Importantly, with the development of an economic infrastructure in space, the demarcation between commerce, science and exploration dissolves. Each is essential to the fulfilment of the other. The intensity of scientific exploration required for industrial expansion in space lends itself to a strong partnership between the scientific, exploration and commercial sectors. For example, lunar and asteroidal assays will use the same tools as those used for scientific research, and determining the composition, distribution, mineralogy, and geological history of planetary and asteroidal materials will be required for both commercial and scientific purposes. In addition, the development of a commercial infrastructure in space will provide increased opportunities for constructing astronomical observatories in space, studying of our near space environment, and a host of other scientific endeavours.

The final talk of the morning session was given by **James Carpenter** (European Space Agency) on "*Lunar resources: Exploration enabling science.*" Many space agencies and private entities are looking to lunar resources as a potentially game changing element in future space exploration architectures, increasing scale and reducing cost. This would increase the opportunities for conducting science in space. However, ISRU is not yet an integral component of the Global Exploration Roadmap (ISECG, 2013), or any other agency-led activity. Before decisions can be made on the inclusion of extraterresrial resources in exploration roadmaps some important questions need to be answered about the viability of different resources and the processes for their extraction and utilisation. The missions and measurements that will be required to answer these questions can only be performed through the engagement and support of the scientific community. An example is ESA's proposed PROSPECT (Package for Resource Observation and *In-Situ* Prospecting for Exploration, Commercial exploitation and Transportation) package, which is designed to fly on future lunar polar landers (e.g. Russia's Luna-27 mission proposed for the early 2020s) with the aim of detecting and characterising lunar polar volatiles. ESA is also conducting pre-Phase-A studies of a small rover (Lunar Volatile Prospector), as well as preparing for collaboration with Russia on a lunar polar sample return mission (Luna-29) for the mid-2020s. Last, but not least, ESA is actively exploring the possible role of small satellites, such as cubesats and penetrators, for fundamental scientific investigations as well as resource prospecting. All such missions will both add to scientific knowledge and answer key questions related to lunar resource availability. In this way, science will enable exploration and exploration will enable science.

After lunch, **Simeon Barber** (Open University) provided more details about PROSPECT with his talk entitled "*PROSPECT: ESA's lunar resource prospecting tool as a precursor to lunar science and exploration.*" He began by noting that although orbital measurements, and the 2009 LCROSS impactor, have provided persuasive evidence for lunar polar ice, many uncertainties remain concerning its distribution and accessibility. The unanswered questions will best be addressed by landing spacecraft in the polar regions and prospecting *in-situ* for lunar water and other volatiles. The key requirements for PROSPECT are informed by, and seek to provide ground-truth measurements to constrain the existing remote-sensing datasets and modelling studies. PROSPECT shall therefore obtain surface samples (for comparison with orbital infra-red data) and sub-surface samples from depths to 1.2 – 2 m (corresponding to depths probed by orbital neutron spectroscopy). The samples will be analysed using the on-board ProSPA (PROSPECT Sample Processing and Analysis) laboratory. Once delivered to ProSPA, the samples will be imaged for geological context and heated in a controlled manner to release their volatile components as a function of temperature. The composition, including the isotopic ratios, of the volatiles will then be determined by mass spectrometry. ProSPA will provide an option to introduce reducing agents into the ovens during heating, in an attempt to study the efficacy of some of the volatile extraction techniques that have been proposed in the context of ISRU (for example ilmenite reduction by hydrogen gas). The data obtained will generate new understanding of lunar polar resources. This in turn will inform the design and targeting of future more advanced lunar polar missions, such as rovers and sample return missions, as well as longer term ambitions for human outposts on the Moon.

The next speaker was **Vibha Srivastava** (Open University) who talked about *"Microwave processing of lunar soil for supporting longer-term exploration missions on the Moon."* Vibha noted that future human space exploration will involve long term stays on the surfaces of other planetary bodies, and that this will necessitate the utilisation of local resources for building a local infrastructure (e.g. buildings, roads, launch pads, etc). In the case of the Moon, lunar soil appears to be an ideal feedstock for building up an infrastructure, but there are significant knowledge gaps in terms of the chemical and physical properties of lunar soil. These knowledge gaps will need to be filled in order to develop appropriate construction techniques for lunar applications. Additive manufacturing (aka 3D printing) is a potentially valuable technology for infrastructure development on the Moon and asteroids (Fig. 3), and Vibha briefly outlined current thinking in this exciting area. One possible 3D printing technique that has been proposed in this context is the microwave processing of lunar soil (e.g. Taylor and Meek, 2005), and understanding the dielectric behaviour of the regolith in the microwave region of the spectrum is increasingly recognized as an important topic of research in the field. In her presentation, Vibha reviewed the current status of our knowledge of the potential of microwave processing of lunar soil for construction purposes on the Moon and outlined future research priorities. In addition to paving the way (perhaps literally!) to the construction of lunar habitats, this work is also expected to benefit the planetary science community by improving our knowledge of the physical characteristics of lunar and planetary regoliths.

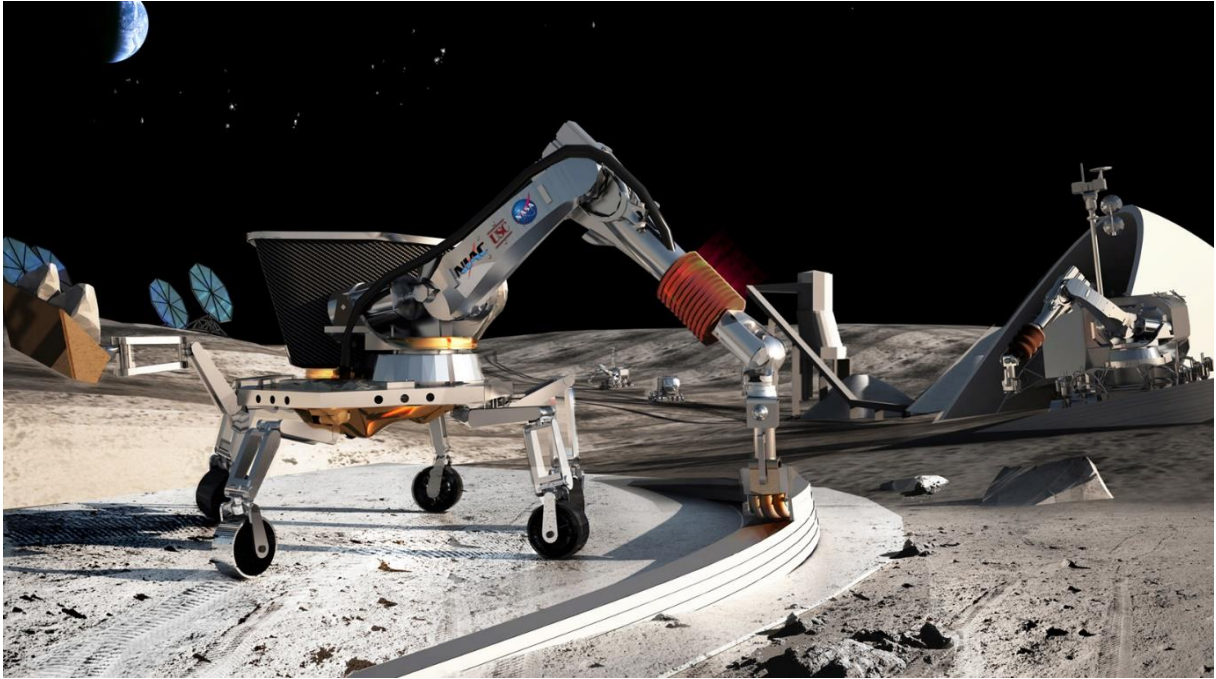

**Fig. 3**. Artist's conception of a robotic 3-D printer constructing a lunar habitat using regolith as a feedstock. Using local materials in this way will enable the construction of scientific and commercial infrastructure on the surfaces of the Moon, Mars, and perhaps elsewhere (NASA).

Moving outwards from the Moon, the next talk was given by **Manuel Grande** (Aberystwyth University) who asked the question "*Asteroid mining: What's in it for planetary science?*" Manuel began by taking us back to the writings of Konstantin Tsiolkovsky (1857-1935), the great pioneer and prophet of astronautics, who had already identified the desirability of utilising asteroidal resources for space development in 1900. He then drew attention to forthcoming scientific missions to asteroids (i.e. Hayabusa 2 and OSIRIS-Rex), as well as proposals to deflect the orbits of small asteroids (e.g. NASA's Asteroid Redirect Mission; Fig. 4) and the international AIDA (Asteroid Impact and Deflection Assessment) mission. All these missions will add to our scientific understanding of asteroids, but will also allow us to assess the extent to which they may contain economically valuable materials. Indeed, the potential economic wealth of asteroids (see http://www.asterank.com/) has led to the creation of a number of private companies seeking to exploit them. Manuel argued that the scientific community stands to gain much from this activity because improved access to a wide range of asteroids will increase opportunities for sample return, *in situ* measurements, and ground-truthing of remote-sensing data. All these activities will increase or understanding of the origin and early evolution of the Solar System. It is therefore urgent that the scientific community starts to think about how best to engage with the emerging asteroid mining industry.

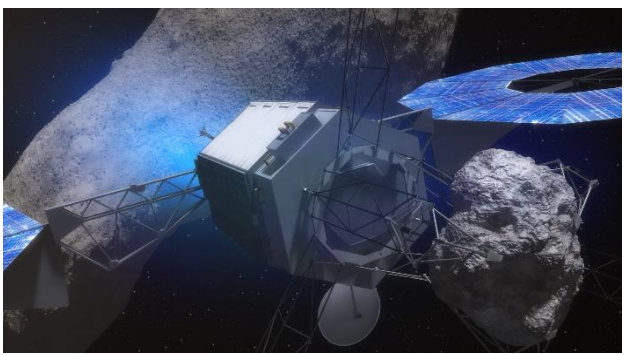

**Fig. 4**. The current version of NASA's proposed Asteroid Redirect Mission (ARM) envisages a spacecraft collecting a small (~6m) boulder from the surface of an asteroid and taking it to lunar orbit for further study. Such a mission would enable planetary scientists to learn a lot about asteroids, but will also help in developing the techniques needed for the utilisation of asteroidal raw materials (NASA).

**Colin Snodgrass** (Open University) then spoke on "*Searching for water in asteroids.*" Colin reiterated the point that water ice is a valuable resource in space, and its presence in small bodies opens the possibility of extracting it directly rather than lifting it from Earth. Comets contain significant ice but are difficult to reach, while asteroids were traditionally thought of as dry rocks. Colin reviewed recent discoveries that show that there is water ice in asteroids. Direct detections have been made for Ceres and other large asteroids, and the presence of water ice is implied in small 'active asteroids' and 'main belt comets.' Confirming these detections, and identifying ice in other asteroids, would benefit from a small (e.g. ESA S-class; < 100M€) survey mission that would fly a modest UV telescope (30-50 cm aperture) above the atmosphere in order to access the 308 nm OH band. Alternatively, such a survey could be conducted by a larger, more general-purpose, UV space telescope that would have wider astrophysical applications. Astronomical surveys of this kind could be complemented by a dedicated spacecraft that would tour the asteroid belt in order to conduct flybys of multiple asteroid targets. The proposed CASTAway concept, currently being studied in the context of a future ESA medium-class mission, performs both a survey and a tour (see http://bit.ly/castawaymission). Cataloguing the abundance of water-rich asteroids is scientifically important, because of what it would tell us about the history and evolution of the Solar System, but would also be key to identifying resource-rich asteroids that may be economically exploitable.

The asteroid theme was continued by **Colin McInnes** (Glasgow University) who spoke on "*Near Earth asteroid capture dynamics, material sorting and utilisation for large on-orbit reflectors.*" Colin outlined a range of possible strategies for capturing small near Earth asteroids and transferring them to Earth or lunar orbit. It can be shown that the energy required for capture can be minimised by using the stable manifolds of periodic orbits in the Sun-Earth and Earth-Moon systems, such that modest velocity changes of less than 500 ms$^{-1}$ are required for the most easily accessible objects (Garcia Yarnoz et al., 2013). Harvesting such easily accessible objects could be undertaken at an energy cost far less than lifting material from the lunar surface. Once in Earth orbit this captured material would be available for scientific study and/or commercial applications. One possibility would be to use solar radiation pressure to sort asteroidal material (e.g. to separate chondritic from denser metallic grains) as a first step towards commercial applications. Colin also showed that metallic asteroidal material thus concentrated could be used to construct large-scale thin-film parabolic reflectors in space that could have a wide range of future applications, for example as telescope mirrors, sunlight concentrators (or shields) and solar sails (Borgraffe et al., 2015).

The penultimate talk was given by **Michael Johnson** (Imperial College London/ PocketSpacecraft.com) on "*Replenishing prepositioned Spacecraft-on-Demand printers.*" Michael began by drawing attention to the success of the cubesat concept, and the increasing roles foreseen for cubesats in scientific and commercial space activities. He argued that still smaller spacecraft will be possible, and that these will reduce the costs of access to space to levels where individuals will be able to afford to buy, launch and operate them with little or no technical expertise. The *in situ* production of space systems of this kind for science, exploration, education and profit is likely to be the preferred method of implementing large-scale space systems in the future. The Spacecraft-on-Demand system is being developed to test approaches for realising spacecraft, landers and rovers with masses in the range of mg to kg. A terrestrial laboratory system is currently being repurposed as a Prepositioned Orbiting Printer CubeSat (POP/C) to demonstrate production of entire functional space systems on orbit. POP/C is designed to launch with a modular detachable Insulator, Conductor, Energy and Semiconductor CubeSat (ICES/C) materials cartridge to demonstrate production of multiple devices per cartridge. Although the initial devices are designed to be replenished by replacing a depleted ICES/C with a replacement launched from Earth, approaches for refilling expended cartridges with suitable material from the environment are also being explored. This would greatly increase the affordability and flexibility of such systems for scientific and commercial applications.

The final talk was given by **Ian Crawford** (Birkbeck College London) on "*The long-term scientific benefits of space infrastructure.*" Ian argued that there are at least four (non-mutually exclusive) scientific benefits of space resources: (i) scientific discoveries made during prospecting for and extraction of space resources; (ii) using space resources to build, provision and maintain scientific instruments and outposts (i.e. ISRU); (iii) leveraging economic wealth generated by commercial space activities to help pay for space science activities (e.g. by taxing profits from asteroid mining, space tourism, etc); and (iv) utilisation of the transportation and other infrastructure developed to support commercial space activities. Concentrating on the latter point, Ian argued that science will be a major beneficiary of space infrastructure, even if its major elements (e.g. spaceships, habitats, mining activities, or in-space construction capabilities; Fig. 5) are not developed with science primarily in mind (e.g. Crawford, 2001). As an example, he drew a parallel with the way in which geologists and astronomers on Earth make use of a civil aviation infrastructure that was mainly developed for tourism in order to get to their field sites and observatories, without having to design, build, and operate commercial airliners. Future scientists operating on the Moon and Mars would similarly benefit from a commercial interplanetary transportation infrastructure. Other examples of scientific activities that would be facilitated by the development of space infrastructure include the construction of large space telescopes, the establishment of scientific research stations on the Moon and Mars, and, in the more distant future, the construction of interstellar space probes for the exploration of planets around nearby stars (e.g. Crawford, 2010).

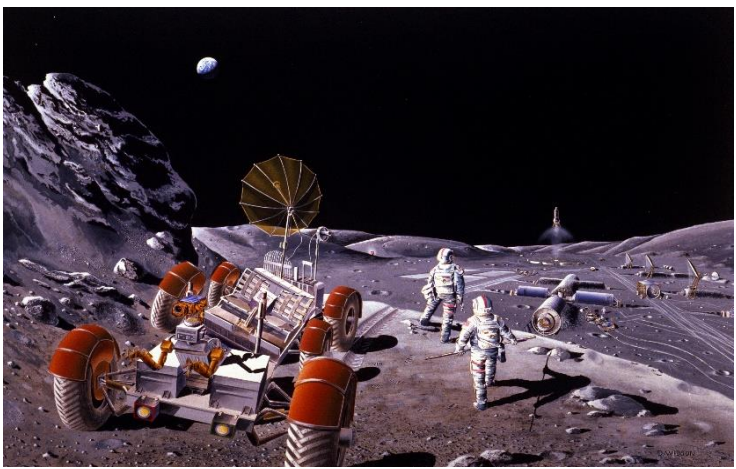

**Fig. 5**. Infrastructure on the Moon. This artist's drawing of future lunar exploration includes multiple infrastructural elements, including habitats, local transportation (rover), interplanetary transportation (rocket), and trained personnel (astronauts). Sustainable operation of such an infrastructure will require the use of *in situ* resources and will enable lunar (and later martian) exploration on a scale that will not otherwise be possible (NASA).

Overall, the meeting, which was attended by over 60 participants, demonstrated widespread interest in the potential scientific benefits of space resource utilisation. It is intended that several of the papers presented at the meeting, together with others on the same theme, will appear in a special issue of the journal *Space Policy*.

**Acknowledgements**

We thank the RAS for supporting this meeting, and the speakers for their contributions which resulted in such an interesting meeting. The summaries of talks given above are based on the abstracts provided by each named speaker, supplemented by our own notes. It is intended that more detailed versions of some of the above contributions will appear in a forthcoming issue of the international peer-reviewed journal *Space Policy*.